\documentclass[a4paper,12pt]{JHEP}
\usepackage{amsmath,amsfonts,amssymb}
\usepackage{graphicx} 

\usepackage{cite}

\numberwithin{equation}{section}

\newdimen\tableauside\tableauside=1.0ex
\newdimen\tableaurule\tableaurule=0.4pt
\newdimen\tableaustep
\def\phantomhrule#1{\hbox{\vbox to0pt{\hrule height%
	\tableaurule width#1\vss}}}
\def\phantomvrule#1{\vbox{\hbox to0pt{\vrule width%
	\tableaurule height#1\hss}}}
\def\sqr{\vbox{%
  \phantomhrule\tableaustep
\hbox{\phantomvrule\tableaustep\kern\tableaustep\phantomvrule\tableaustep}%

\hbox{\vbox{\phantomhrule\tableauside}\kern-\tableaurule}}}
\def\squares#1{\hbox{\count0=#1\noindent\loop\sqr
  \advance\count0 by-1 \ifnum\count0>0\repeat}}
\def\tableau#1{\vcenter{\offinterlineskip
  \tableaustep=\tableauside\advance\tableaustep by-\tableaurule
  \kern\normallineskip\hbox
    {\kern\normallineskip\vbox
      {\gettableau#1 0 }%
     \kern\normallineskip\kern\tableaurule}%
  \kern\normallineskip\kern\tableaurule}}
\def\gettableau#1 {\ifnum#1=0\let\next=\null\else
  \squares{#1}\let\next=\gettableau\fi\next}

\def\cL{{\cal L}}

\def\cO{\cal O}

\newfont{\goth}{eufm10 scaled \magstep1}

\def\a{\alpha}

\def\c{\gamma}
\def\d{\delta}
\def\e{\epsilon}
\def\f{\phi}
\def\vf{\varphi}

\def\la{\lambda}\def\L{\Lambda}

\def\th{\theta}

\def\del{\partial}

\def\ft{\widetilde{\phi}}

\def \ys {{y\kern-.5em / \kern.3em}}

\def\bd{\begin{document}}
\def\ed{\end{document}}
\def\ba{\begin{array}}
\def\ea{\end{array}}
\def\bea{\begin{eqnarray}}
\def\eea{\end{eqnarray}}
\def\fft#1#2{{#1 \over #2}}
\def\be{\begin{equation}}
\def\ee{\end{equation}}
\newcommand{\eq}[1]{(\ref{#1})}
\def\eqs#1#2{(\ref{#1}-\ref{#2})}
\def\det{{\rm det\,}}
\def\tr{{\rm tr}}
\newcommand{\ho}[1]{$\, ^{#1}$}
\newcommand{\hoch}[1]{$\, ^{#1}$}
\def\ra{\rightarrow}
\def\uha{{\hat {\underline{\a}} }}
\def\uhc{{\hat {\underline{\c}} }}

\def \Om {\Omega}
\def \bfd {{\bf d}}
\def \del {\partial}
\def \eps {\epsilon}
\def \Z {{\bf Z}}
\def \xb {\bar{x}}
\def \lg {\langle}
\def \rg {\rangle}
\def \Omt {\tilde \Omega} 
\def\l{\left}
\def\r{\right}
\def\kt{\widetilde{k}}
\def\Kt{\widetilde{K}}
\def\co{\rm cos}
\def\si{\rm sin}
\def\vft{\widetilde{\varphi}}

\title{
On the Operator Product Expansion
in Noncommutative Quantum Field Theory}
\author{\hspace{4mm} Frederic Zamora\\
    Theory Group, Physics Department\\
    University of Texas at Austin\\
    Austin TX 78712 USA.\\
{~}\\
\email{zamora@zerbina.ph.utexas.edu}
}
\abstract{
Motivated by the mixing of UV and IR effects,
we test the OPE formula in noncommutative field theory.
First we look at the renormalization of local
composite operators, identifying some of their 
characteristic IR/UV singularities.
Then we find that the product of two fields
in general cannot be described by a series expansion
of single local operator insertions.
}
\preprint{
UTTG--06-00\\
{\tt hep-th/0004085}\\
April 11, 2000
}

\begin{document}

\section{Introduction}\label{sec:intro}

To send the location of two operators 
to the same spacetime point is a singular 
process in local Quantum Field Theory (QFT) if the 
ultraviolet (UV) cut-off mass is not finite. 
This combination of locality
with UV divergences is represented 
by the Operator Product Expansion (OPE) \cite{Wilson:1969zs}, 
where the short distance dependence 
between the locations of the two operators is encoded 
in the Wilson coefficients. 
The very nature of the OPE is much in the spirit 
of the Wilson renormalization group approach to field theories: 
the decoupling of scales allows us to codify the 
short distance effects in the coeffients multiplying
a series of insertions of local composite operators.
Another way to see the divergences associated 
with the Wilson coefficients is to observe that
sending the cut-off to infinity and locating the operators 
at the same spacetime point are two different limits 
which in general do not commute.

Recently, Noncommutative Quantum Field Theory (NCQFT)
has been the subject of an intense research,
either by using field theory methods 
\cite{Filk:1996dm,Martin:1999aq,
Sheikh-Jabbari:1999iw,Krajewski:1999ja,Chepelev:1999tt,
Minwalla:1999px,VanRaamsdonk:2000rr,
Hayakawa:1999zf,Matusis:2000jf,Grosse:2000yy,
Aref'eva:1999sn,Aref'eva:2000hq,Aref'eva:2000bg,
Arcioni:1999hw,Fischler:2000fv,Fischler:2000bp,
Ardalan:2000cy,Gracia-Bondia:2000pz,
Campbell:2000ug,Bonora:2000he,Chu:2000bz,
Ambjorn:2000nb,Gopakumar:2000zd}
or by exploting its embeding into string theory
\cite{Connes:1998cr,Schomerus:1999ug,Chu:1998qz,Chu:1999ij,
Seiberg:1999vs,Sheikh-Jabbari:1999vm,Bigatti:1999iz,
Ishibashi:1999hs,Iso:2000ew,
Hashimoto:1999ut,Maldacena:1999mh,Alishahiha:1999ci,
Barbon:1999mx,Sheikh-Jabbari:2000en,Mateos:2000qq,
Andreev:2000rm,Kiem:2000wt,Bilal:2000bk,Gomis:2000bn,
Rajaraman:2000dw,Liu:2000qh}.
Up to now, one of its most intriguing features
is the mixing of UV with IR effects \cite{Minwalla:1999px}.
Since the OPE is such an important 
and characteristic property of local QFT, it is very 
natural to test it in NCQFT.
Due to the UV/IR mixing of scales, 
something different should happen 
in the process of sending operators 
to the same spacetime point.

The simplest non-trivial framework to analyze this
is via a single scalar with a cubic interaction,
\begin{equation}
{\cL} = \frac{1}{2}(\del\f)\star(\del\f) 
+ \frac{1}{2}m^2\f\star\f 
+\frac{g}{6}\f\star\f\star\f \quad.
\end{equation}
We will work in six dimensions, with $\th^{0i}=0$,
and Wick-rotate to the Euclidian signature. 
Our analysis will be perturbative and mainly restricted to one-loop.
But we expect our results to hold at higher loops,
provided that the renormalization program 
\`a la Dyson (without keeping a finite cut-off)
can be extended at a multi-loop level. 

In section 2 we analyze the insertion of single renormalized
composite operators in noncommutative scalar field theories,
focusing on their IR/UV singularities.
In section 3 we look for singularities related to 
the product of two fundamental fields in the six dimensional
$\f^3$ field theory. We see that $\th^{ij}\not=0$ modifies 
the analytic structure of the Green functions, 
preventing the possibility of replacing the product of 
two fundamental fields by a series expansion of single
local composite operator insertions.
The possible stringy explanations of this work is left
for the future.

\bigskip

\section{Renormalization of Composite Operators in NCQFT}

\subsection*{UV divergences}

Before testing the OPE in NCQFT, we should first look at 
the properties of the local composite operators.
By local composite operator we understand an arbitrary
number of fields and/or derivatives of fields,
all evaluated at the same spacetime point.
In local QFT, 
the insertion of a bare composite operator ${\cal O}_0$ 
in a general n-point Green function carries UV divergences 
which are intrinsic to the composite operator
and require their own renormalization. The symbol
for the renormalized composite operator, {\it i.e.}, the one 
whose insertion in a general n-point Green function is finite 
at infinite cut-off, is expressed by $ [\cal O]$,
to distinguish from $\cO$, the simple product of 
renormalized fields, which is a different object.

Furthermore, the renormalization produces 
operator mixing,
\begin{equation}
[{\cal O}_i](\mu) = \sum_{j;\ d_j \leq d_i}Z_{ij}\l(\L/\mu;g(\mu)\r)
\L^{d_i-d_j}{\cal O}_{0,j}(\L) \quad ,
\end{equation}
where the sum only involves
operators whose canonical mass dimension $d_j$ is smaller 
or equal than $d_i$.
$\L$ is the UV cut-off, $\mu$ the renormalization scale
and $g(\mu)$ the renormalized coupling constant.

In principle any local composite operator is a 
linear combination of $\{\f^2, 
(\del^2\f)\f, \cdots \}$. But it is very likely that
in NCQFT it is more convenient to use the basis 
of operators where
any product of fields is accomplished by the Moyal product:
$\{\f\star\f,(\del^2\f)\star\f, \cdots\}$.
Then, a very natural starting point
is to look at the renormalization of $(\f\star\f)(x)$.
The Fourier transform in momentum space of the bare operator is
\begin{equation}
\widetilde{(\f_0\star\f_0)}(p) =\int \frac{d^D x}{(2\pi)^D}\ 
e^{-ipx} (\f_0\star\f_0)(x) 
=\int \frac{d^D q}{(2\pi)^D}\ 
e^{ip\wedge q} \ft_0(q)\ft_0(p-q) \quad.
\end{equation}
with $p\wedge q \equiv p^i\th^{ij}q^j$.
The insertion of this operator at tree level is
\begin{equation}
\raisebox{-9ex}{\includegraphics[width=14ex]{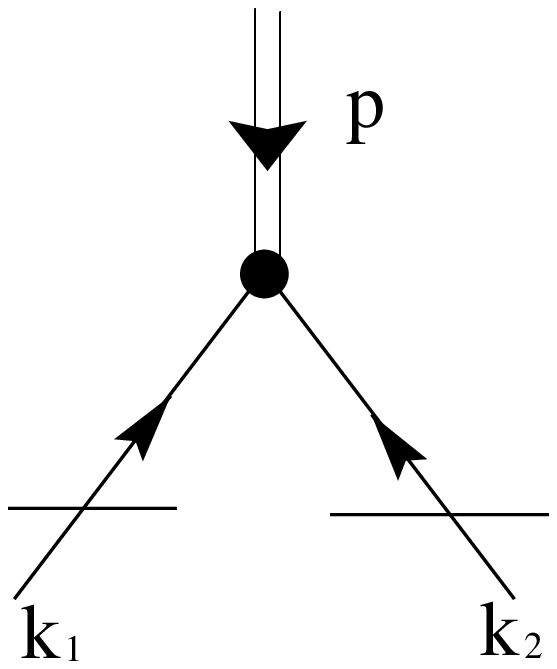}}
\hspace{8mm}=(2\pi)^D \d^D(p+k_1+k_2)\ 2 {\co}(k_1\wedge k_2) \quad.
\label{tree}
\end{equation}
where the horizontal lines in the $k_1$ and $k_2$ momentum legs
mean that their propagators have been amputated.

For a general diagram with superficial degree of divergence 
$\omega$, the insertion of $\f\star\f$ decreases it to $\omega-2$.
In the six dimensional $\f^3$ field theory, 
the superficial degree of divergence of an $n$-point function
is $w=2(n-3)$.
Therefore, to analyze the renormalization of $\f\star\f$,
we just have to consider its insertion at $n=1,2$.
Since we only look at the connected Green functions
with (amputated) external legs, we ignore the $n=0$ case
\footnote{If not, the identity operator enters into 
the renormalization.}.

\begin{figure}
\begin{minipage}[b]{.35\linewidth}\centering
\mbox{\includegraphics[width=1.8in]{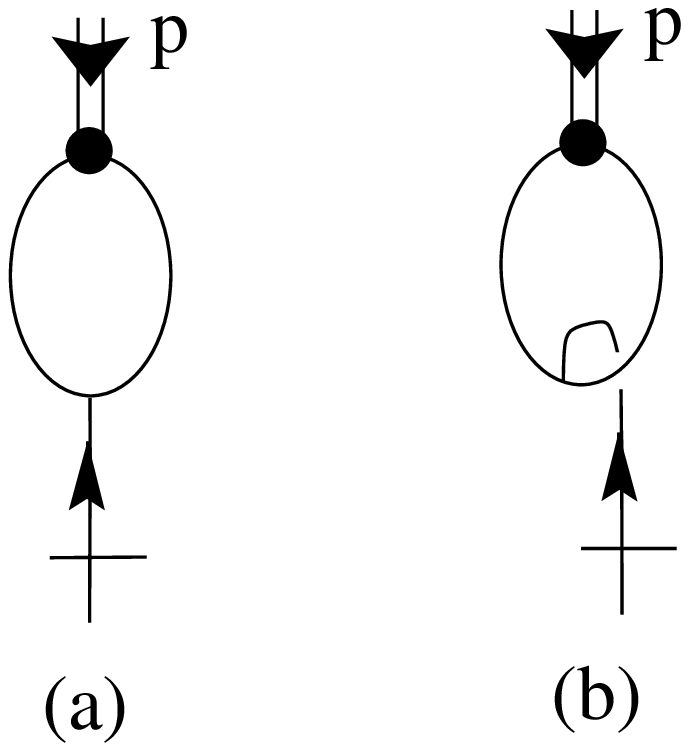}}
\caption{One-loop insertion in 1PI one-point function.}
\label{fig:1point}
\end{minipage}\hfill
\begin{minipage}[b]{.58\linewidth}\centering
\mbox{\includegraphics[width=3.8in]{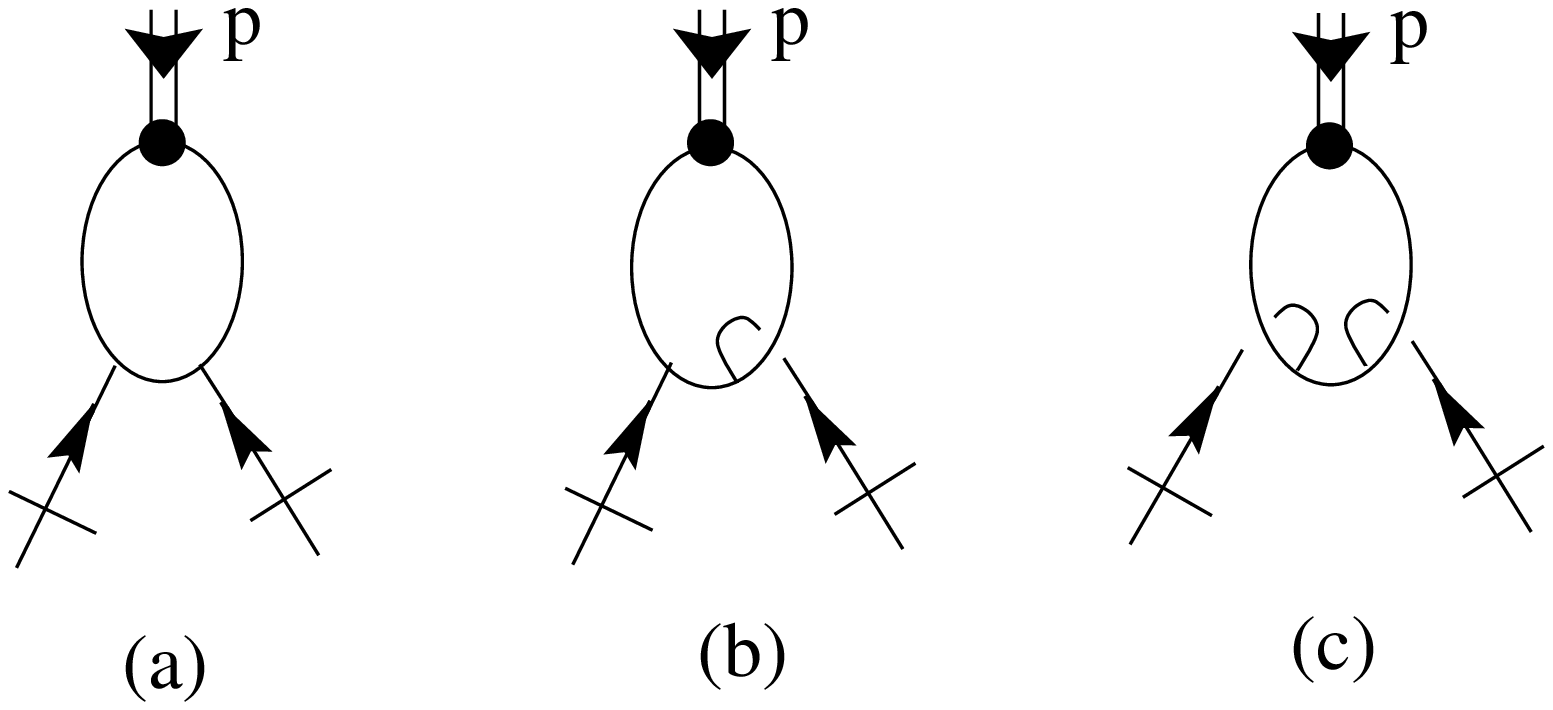}}
\caption{One-loop insertion in 1PI two-point function.}
\label{fig:2point}
\end{minipage}
\end{figure}

The type of one-loop insertions in the 1PI one and two
point functions are shown in figures \ref{fig:1point} and 
\ref{fig:2point}. 
The UV divergences \footnote{We use the Schwinger cut-off regularization 
\cite{Minwalla:1999px}.}
of the planar graphs
can be subtracted by the introduction of the renormalized composite
operator
\begin{equation}
\begin{split}
[\f\star\f] = &\frac{g}{2^5\pi^3}\L^2\f_0 - \frac{g}{2^6\pi^3}
\ln\l(\frac{\L}{\mu}\r)\l(\frac{\del^2}{6}+m^2\r)\f_0 
\\
&+\l(1-\frac{g^2}{2^8\pi^3}\ln\l(\frac{\L}{\mu}\r)\r)\f_0\star\f_0 
+{\cO}(g^3)\quad .
\end{split}
\end{equation}
The planar graph in figure \ref{fig:1point} (a) produces
the operator mixing with $\f_0$.
To cancel the UV divergence of the graph in
figure \ref{fig:2point} (a) one has to introduce a counter-term
for $\f_0\star\f_0$. Observe that 
the renormalization of $\f\star\f$ preserves 
the $\star$-product structure.
We could look at the renormalization of $[\f\star\f\star\f]$
and obtain the same structure:
it mixes with $\f_0$, $\f_0\star\f_0$ and $\f_0\star\f_0\star\f_0$,
because of the UV planar graphs with one, two and three legs 
(respectively) attached to the loop.
The general result is that 
the basis of composite operators with the $\star$-product
structure is closed under renormalization.
We think that this fact is crucially related to the (possible)
renormalizability of the noncommutative field theory.

\subsection*{Nonplanar Graphs}

But the planar graphs are not the full story.
For the one-point function, the non-planar contribution 
of figure \ref{fig:1point} (b) is 
\begin{equation}
\begin{split}
\lg[\widetilde{\f\star\f}&](p)\ft(k)\rg_{NP} =  - \frac{g}{2} 
(2\pi)^6 \d^6(p+k)
\int \frac{d^6 p}{(2\pi)^6}\frac{e^{ip\wedge k}}
{\l(p^2 +m^2 \r)\l( (p+k)^2 +m^2 \r)} 
\\
\sim & - \frac{g}{16\pi^3} (2\pi)^6 \d^6(p+k)
\l( \frac{2}{|\kt|^2} +
\l(\frac{k^2}{6}+m^2\r)\ln(|\kt|\mu) +\cdots \r)
\ {\rm when}\ |\kt|\simeq 0 \ .
\end{split}
\end{equation}
The exact result for the integral can be obtained 
from $I_2(k;\kt)$ in the appendix. 
If the insertion of the composite operator is 
at zero momentum in the noncommutative directions,
it appears precisely at the IR singularity 
$\kt^i \equiv\theta^{ij}k^j=0$.

The non-planar contribution to the 
amputated and connected two-point function is
given by the diagrams in fig. \ref{fig:1point} (b) and 
fig. \ref{fig:2point} (b) and (c)
\footnote{The functions $I_2[p;\kt]$ and $J_0[p_1,p_2;\kt]$
are defined in the appendix.},
\begin{equation}
\begin{split}
\lg[\widetilde{\f\star\f}](p) \ft(k_1)\ft(k_2)\rg_{NP} 
& = \frac{g^2}{2} (2\pi)^6 \d^6(p+k_1+k_2)
\\
&\cdot\l\{{\co}(k_1\wedge k_2)
\l(\frac{1}{p^2+m^2}I_2[p;\tilde{p}]
+J_0[p,k_1;\tilde{p}]\r)\r.
\\
&\ \l. +e^{ik_1\wedge k_2} J_0[p,k_1;2\kt_1]
+e^{-ik_1\wedge k_2} J_0[p,k_2;2\kt_2]\r\} \quad .
\label{n2p6D}
\end{split}
\end{equation}
Again, for $\tilde p=0$, its insertion is singular. 
This situation generalizes to any operator insertion
at zero momentum in the noncommutative directions.
The singularities appear because of the 
high-momentum circulating in the 
non-planar loop. Therefore, also for composite operators 
there is the UV/IR mixing noticed in \cite{Minwalla:1999px}.
In this case, it is caused by the UV divergences 
associated with the composite operators. 

Another way to understand this singularity is to 
observe that, contrary to the case of local QFT,
a composite local operator ${\cO}_i(x)$ and its 
corresponding coupling parameter $g_i$
have \emph{different} multiplicative renormalizations,
with the difference parametrized precisely 
by the IR divergence of ${\cal\widetilde O}_i(p)$ at $\widetilde p=0$
If one restores the cut-off and then takes 
${\widetilde p}=0$, one recovers the usual property
\begin{equation}
\int dx_{nc}\ g_{i,0}(\L){\cO}_{i,\L}(x) 
= \int dx_{nc}\ g_i(\mu) [{\cO}_i]_\mu(x) 
\quad +{\rm lower\ dim.\ ops.}
\end{equation}
So, it is crucial that the parameter $g_i$ 
\emph{only} couples to the spacetime integral of ${\cO}_i$.

Besides this intrinsic IR singularity for the zero momentum
insertion of $\widetilde{\cO}$, 
a general $n$-point function will have a multiple 
set of singularities located at different linear 
combinations of the external momenta, $\sum_i a_i \kt_i =0$,
with the constants $a_i$  related to the different 
momentum channels or graphs. For instance, in 
\eqref{n2p6D} there are additional one-loop singularities 
located at $\kt_1=0$ and $\kt_2=0$.

\subsection*{Composite Operators with some $\star$-products missing}

Even though they are odd objects in NCQFT, one could ask
about the quantum properties of composite operators
which lack some $\star$-products, like $[\f^2]$, 
$[(\f\star\f)\f]$, etc. Formally, they could be expressed
as an infinite sum of operators with 
every product given by the $\star$-product,
which would correspond to working 
in the $\star$-product operator basis.
But at the perturbative level we can look at, for instance,
$[\f^2](x)$ as a symbol, with its tree level insertion 
just given by \eqref{tree} without the phase factor.
In this case, the one-loop integrals corresponding 
to figures 1 and 2 are all finite,
since they always have a Moyal phase, 
coming from the interaction vertices, which cuts off the high
frequency modes in the loop\footnote{We have verified 
this situation holds up to two loops.}.
Then, at one-loop, $[\f^2]$ only renormalizes with 
the identity operator.
We can also look at $[\f^3]$,
defined by its obvious insertion 
in the three point function at zero order in the coupling constant. 
In this case, it is only UV divergent when inserted in the
one-point function. It can be made finite simply by 
adding a counter-term proportional to $\f_0$.
Finally, for the insertion of $[(\f\star\f)\f]$,
the necessary counter-terms to make it one-loop
finite are $\f_0$, $\f_0^2$ and $(\f_0\star\f_0)\f_0$.

Since the UV properties of these operators
are different from the analogous operators
with all the products given by the $\star$-product,
we expect that the location 
of their singularities associated to the $\th\to 0$ limit
to be also different.
Indeed, in general their insertion at zero momentum 
is still divergent
\footnote{which in the case of $[\f^2]$ is
fully related to the UV divergence of the mass parameter.}.
But when inserted in a general $n$-point function, 
their additional singularities are
located at different places from the ones corresponding 
to the insertions of the $\star$-product's operators.
For instance, the insertion of $[\widetilde{\f^2}]$
in the two-point function produces singularities at
$\kt_1\pm\kt_2 =0$.

\bigskip

\subsection*{Noncommutative $\la\varphi^4$ in Four Dimensions}

Same qualitative results are obtained for the case 
of a four dimensional $\la\varphi^4$ interaction,
\begin{equation}
{\cL} = \frac{1}{2}(\del\varphi)\star(\del\varphi)
+ \frac{1}{2}m^2\varphi\star\varphi
+\frac{\la}{4!}\varphi\star\varphi\star\varphi\star\varphi \quad.
\end{equation}
In this case, the discrete symmetry $\varphi\to -\varphi$  
prevents $\varphi\star\varphi$ from mixing with $\varphi$.
Looking at its insertion into the (amputated and connected)  
two-point function at one-loop, we get the renormalization
\begin{equation}
[\varphi\star\varphi] = \l(1-\frac{\la}{2^4 3\pi^2}
\ln\l(\frac{\L}{\mu}\r)\r)\varphi_0\star\varphi_0 \quad.
\label{2vf}
\end{equation}
It does not correspond to the inverse 
multiplicative renormalization of the mass squared,
which is
\begin{equation}
m^2=\l(1+\frac{\la}{2^3 3\pi^2}\ln\l(\frac{\L}{\mu}\r)\r)
m^2_0 -\frac{\la}{12\pi^2}\L^2 \quad.
\label{mass}
\end{equation}
As explained before, the reason is a singularity for 
the insertion of $[\widetilde{\varphi\star\varphi}](p)$ 
at zero momentum in the noncommutative directions.
For $\widetilde{p}\simeq 0$, 
\begin{equation}
\begin{split}
\lg[\frac{1}{2}\widetilde{\vf\star\vf}](p)\vft(k_1)\vft(k_2)\rg_{NP} 
\sim & (2\pi)^4\d^4(p+k_1+k_2)
{\co}(k_1\wedge k_2)
\\
&\cdot\l(-\frac{\la}{2^4 3\pi^2}\r)
\l(\frac{2}{|\widetilde{p}|} +
\ln(|\widetilde{p}|\mu)\r) \ ,
\end{split}
\end{equation}
whose coefficient of the log divergence 
added to the coefficient of the UV log divergence 
in \eqref{2vf} exactly matches with
the negative of the log coefficient in \eqref{mass}.

We also point out that, 
as in the six dimensional $\f^3$ field theory,
the additional one-loop singularities 
of $[\varphi\star\vf]$ are located at $\kt_1,\kt_2=0$,
and of $[\vf^2]$ at $\kt_1-\kt_2=0$.
The one-loop insertion of the bare operator $\varphi_0^2$ 
in the connected two-point function is also finite.

\bigskip

\section{Operator Product Expansion in NCQFT}

\subsection*{OPE in local QFT}

In NCQFT, the noncommutativity scale $\th\not=0$ 
changes the UV properties of the theory.
As a consequence, renormalized quantities 
do not have a smooth $\th\to 0$ limit; or in other words, 
the two limits $\L\to\infty$ and $\th\to 0$ do not commute
\cite{Minwalla:1999px}.

There is a similar situation in local QFT:
the product of two renormalized operators 
located at different spacetime points, 
$[{\cO}_1](x)[{\cO}_2](y)$ is
singular for $x\to y$ if the UV divergences of $[{\cO}_1{\cO}_2](x)$
are different from the divergences of $[{\cO}_1](x)[{\cO}_2](y)$. 
In this case, the physical meaning for the 
noncommutativity of the $x\to y$ and $\L\to\infty$
limits is encoded in the 
Operator Product Expansion formula:
\begin{equation}
[{\cO}_1]_\mu(x)[{\cO}_2]_\mu(y) = \sum_{n=0}^{\infty} |y-x|^{d_n-d_1-d_2}
C_{12}{}^n\l(|y-x|\mu;g(\mu)\r) [{\cO}_n]_\mu(x) \quad ,
\label{OPE}
\end{equation}
where $\{{\cO}_n \}$ is a convenient basis of local 
composite operators, with canonical mass dimension $d_n$, and 
$\mu$ is the renormalization scale.
The Wilson coefficients $C_{12}{}^n(|y-x|)$ 
can be computed perturbatively, finding that in general 
they are logarithmically divergent when $|y-x|\to 0$
(for field theories defined at a Gaussian fixed point).
For instance, in the case of  
commutative $g\f^3$ scalar theory in six dimensions,
the product of two fundamental fields is
\begin{equation}
\begin{split}
\f\l(x-\frac{\e}{2}\r)\f\l(x+\frac{\e}{2}\r) & = 
\frac{1}{|\e|^2}C_\f(|\e|\mu;g)\phi(x) 
+ C_{\del^2\f}(|\e|\mu;g)\del^2\f(x)
\\
&+ C_{m^2\f}(|\e|\mu;g)m^2\f(x) 
+ C_{[\f^2]}(|\e|\mu;g)[\f^2](x) 
\label{ope6d}
\\
&+ \e^\mu\e^\nu C_{[\del^2\f^2]}(|\e|\mu;g)
[\f\del_\mu\del_\nu\f -\del_\mu\f\del_\nu\f](x) + {\cO}(\e^4) \quad,
\end{split}
\end{equation}
where a perturbative calculation gives the Wilson coefficients
\begin{subequations}
\begin{align}
C_\f(|\e|\mu;g) &= -\frac{g}{16\pi^3} +{\cO}(g^3)
\\
C_{\del^2\f}(|\e|\mu;g)&= \frac{g}{2^7 3\pi^3}\ln(|\e|\mu) +{\cO}(g^3)
\\
C_{m^2\f}(|\e|\mu;g)&= \frac{g}{2^6\pi^3}\ln(|\e|\mu) +{\cO}(g^3)
\\
C_{[\f^2]}(|\e|\mu;g)&=1 -\frac{g^2}{2^6\pi^3}\ln(|\e|\mu)+ {\cO}(g^4) \quad.
\end{align}
\end{subequations}

We want to revisit the same process in NCQFT: 
to analyze the possible singularities associated with
the product of two fields
and see if there is still an Operator
Product Expansion parametrizing it.

\subsection*{Singularities in Position Space}

As stated in the introduction, 
we will limit our analysis to one-loop 
and mainly use the six dimensional $\f^3$ 
field theory as illustrative example.
Fortunately, the results are already non trivial enough
to derive some conclusions.
We will consider an $2+n$-point Green function
as a function of the distance $\e=x-y$ between 
the position of the fields $\f(x)$ and $\f(y)$. 
The rest of the fields in the Green function will
be Fourier transformed to momentum space,
with their external propagators amputated.

We can start with the connected three point function.
Its lowest order contribution can be easily computed
\begin{equation}
\begin{split}
&\lg\f\l(x-\frac{\e}{2}\r)\f\l(x+\frac{\e}{2}\r)\ft (k)\rg 
= -\frac{g}{2}e^{-ik(x-\frac{\e}{2})}
\int \frac{d^6 p}{(2\pi)^6}\frac{e^{ip\e}
\l(e^{ip\wedge k}+e^{-ip\wedge k}\r)}
{\l(p^2 +m^2 \r)\l( (p+k)^2 +m^2 \r)} =
\\
& =  -g e^{-ik(x-\frac{\e}{2})}
\int_0^1 \frac{d\a}{16\pi^2}\ 
\l( \frac{\Delta}{2\pi |\kt+\e|} K_1\l[|\kt +\e|\Delta\r] + 
\frac{\Delta}{2\pi |\kt-\e|} K_1\l[|\kt -\e|\Delta\r] \r)
\label{1p6D}\quad,
\end{split}
\end{equation}
where $\Delta^2 = k^2 \a(1-\a) +m^2$, $\kt^i = \th^{ij}k^j$ 
and $K_n[z]$ is the second kind Bessel function of order $n$.
Most of the results in this section already appear in this 
simple example. 
First, notice that \eqref{1p6D}
is finite for $\kt \pm \e \not=0$. When $\kt \pm \e \simeq 0$,
we have
\begin{equation}
\lg\f\l(x-\frac{\e}{2}\r)\f\l(x+\frac{\e}{2}\r)\ft (k)\rg 
\sim -\frac{g}{64\pi^3} e^{-ik(x-\frac{\e}{2})}
\l( \frac{2}{|\kt\pm \e|^2} +
\l(\frac{k^2}{6}+m^2\r) \ln(|\kt\pm\e|\mu) \r) \quad.
\label{3p6D}
\end{equation}
Again, the dimensionfull scale $\th\not=0$
mixes UV and IR effects. Sending first $\e\to 0$
and then $\kt\to 0$, the divergence in \eqref{3p6D}
is interpreted as IR. If we reverse the order of the 
limits the same divergence has a UV (short distance effect)
interpretation.

In fact, \emph{the length scale $\e$ is replaced 
by the combination $\e\pm\kt$}. 
The product of two fields $\f$, instead
of being singular when they are evaluated at the same spacetime 
point (the situation of local QFT),  
is now singular when the distance between 
them is proportional to the momentum in the noncommutative
directions of the additional external field.
This result supports the picture 
of having extended objects whose characteristic size
is proportional to its momentum
\cite{Sheikh-Jabbari:1999vm,Bigatti:1999iz,
Ishibashi:1999hs,Iso:2000ew}.

We repeat the same calculations for 
the case of the four-point function at one-loop. We get
\begin{equation}
\begin{split}
&\lg\f\l(x-\frac{\e}{2}\r)\f\l(x+\frac{\e}{2}\r)
\ft (k_1)\ft (k_2)\rg 
\\
&=-\frac{g^2}{4}e^{-iK_+(x-\frac{\e}{2})}
\l\{\frac{2{\co}(k_1\wedge k_2)}{K_+^2+m^2}
\l(I_2[K_+;\Kt_++\e]+I_2[K_+;\Kt_+-\e]\r)\r.
\\
&+e^{ik_1\wedge k_2}\l(J_0[K_+,k_1;\Kt_++\e] 
+J_0[K_+,k_2;\Kt_+-\e] \r)
\\
&+e^{-ik_1\wedge k_2}\l(J_0[K_+,k_2;\Kt_++\e] 
+J_0[K_+,k_1;\Kt_+-\e] \r)
\\
&+e^{ik_1\wedge k_2}\l(J_0[K_+,k_1;\Kt_-+\e] 
+J_0[K_+,k_2;\Kt_--\e] \r)
\\
&\l.+e^{-ik_1\wedge k_2}\l(J_0[K_+,k_2;\Kt_-+\e] 
+J_0[K_+,k_1;\Kt_--\e] \r)\r\} \quad,
\label{2p6D}
\end{split}
\end{equation}
where $K_\pm =k_1\pm k_2$. As before, the expression
is finite, unless $\Kt_\pm \pm\e =0$. 

In local QFT, the singularity associated with the product 
of two local operators always appears at the invariant point
$\e=0$. It allows one to define an OPE formula,
where the $\e=0$ singularity can be encoded in 
the universal Wilson coefficients. 
On the contrary, we just saw that in NCQFT the length
scale $\e$ is mixed, via the noncommutativity scale,
with the momenta flowing into the Green function.
In general, for each graph of a given $2+n$-point 
correlation function with fixed momenta $k_i$ 
for the $n$ external legs,
the previous local singularity equation
$\e=0$ is replaced by $\e +\sum a_i \kt_i =0$,
with coefficients $a_i$ depending on the particular graph.
This momentum-dependent shift has 
dramatic consequences for the old OPE formula.
Now, the effects of the short distance scale $\e$
cannot be decoupled and codified in some Wilson
coefficients in front of 
the insertion of single local operators.

Notice that the Taylor expansion in $\e$
for the finite expression on the right-hand side of 
\eqref{1p6D} and \eqref{2p6D}
is obviously reproduced by the corresponding $\e$
expansion of $\f(x-\e/2)\f(x+\e/2)$ on the left-hand side.
But this is far from proving the existence 
of an OPE formula. In particular, the UV/IR singularities
of the composite operators
and an inconsistent RG $\mu$ dependence, already at one-loop,
for the renormalized fields $\f$ and $[\f^2]$, invalidates the 
naive $\e$ expansion for $\f(x-\e/2)\f(x+\e/2)$.
Wilson's OPE has more to do with the relation 
between the analytic properties of the Green functions
and the decoupling of the UV scales.
In fact, the failure of the OPE in NCQFT 
is easily seen in momentum space,
due to the simple expression of the $\star$-product
in that representation.

\subsection*{OPE in Momentum Space}

One can Fourier transform \eqref{OPE} to momentum space 
\footnote{even though in this process the constant terms 
in $C_{12}{}^n (|x-y|)$ are lost unless $q=0$.} 
\begin{equation}
\widetilde{\cO}_1(q)\widetilde{\cO}_2(p-q)
= \sum_n \widetilde{C}_{12}{}^n(q)\widetilde{\cO}_n(p) \quad.
\label{OPE2}
\end{equation}
The decoupling of the scales $q$ and $p$ in this formula
is intimately related to the Wilson renormalization group
approach to field theory. 
The Fourier transformed Wilson coefficients
scale as $|q|^{d_n-d_1-d_2}$, making the OPE 
extremely useful in the regime of $|q|\to\infty$,
where only the lowest dimensional operators are the most relevant ones.
As we know, for generic $q$ the expansion \eqref{OPE2} can be
full of renormalon singularities which spoil its Borel summability.
But the $|q|\to\infty$ limit makes it legitimate to use the OPE.

How much of this still holds in NCQFT? 
Consider a general connected $n+2$-point function in momentum 
representation. Take the momentum of one leg to be $q$ and 
another one $p-q$ and consider the regime where $|q|$
is much larger than any mass scale in the Green function.
One possibility is that the two external legs meet at the 
same interaction vertex. In this case we have 
(the high momentum $q$ flows through the thicker lines) 
\begin{equation}
\begin{split}
\hspace{2cm}
&{\centering\includegraphics[width=35ex]{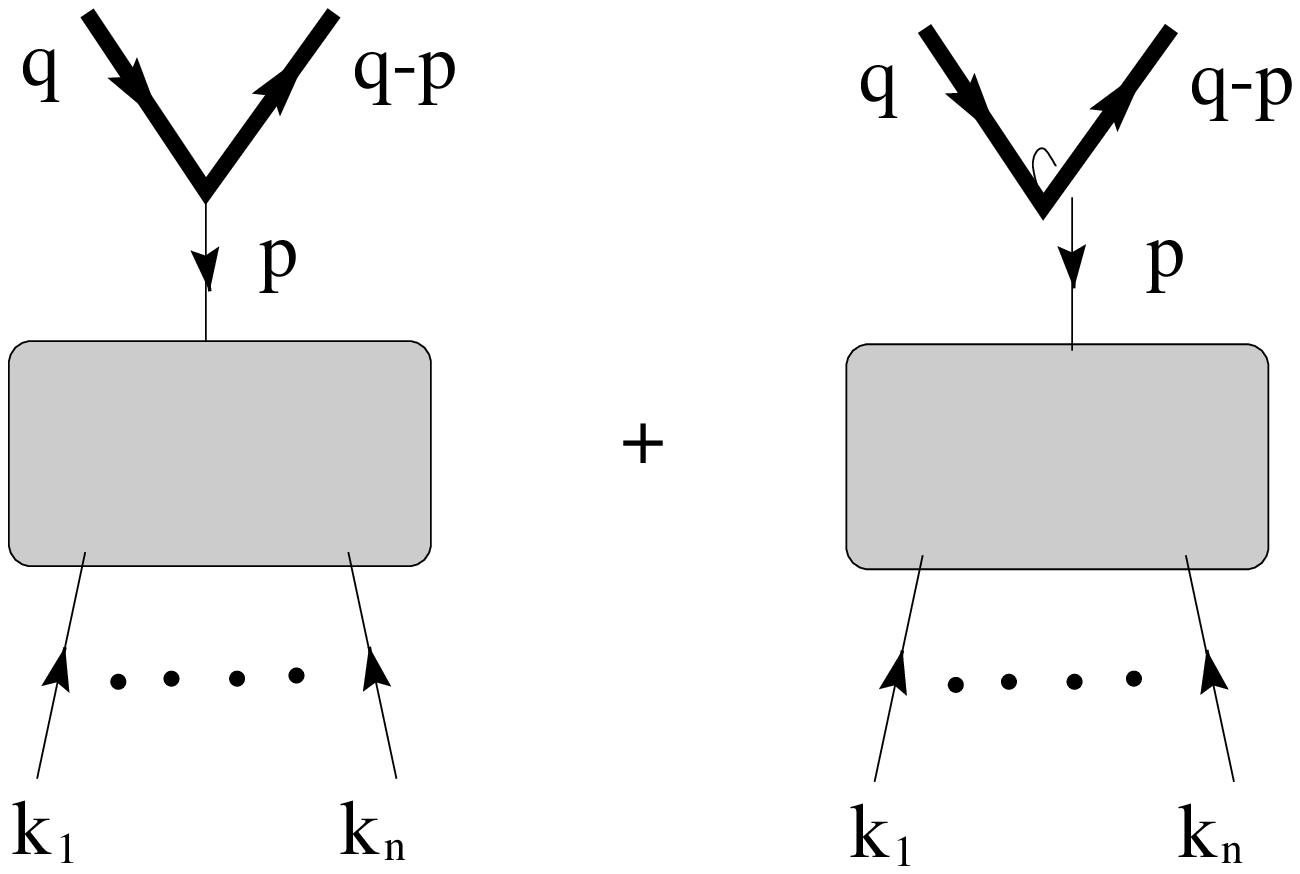}} 
\\
= &{\co}(q\wedge p)
\l(\frac{1}{q^2+m^2}\r)\l(\frac{1}{(q-p)^2+m^2}\r)
G_{n+1}(p,k_1,...,k_2)
\\
= &{\co}(q\wedge p)\l( \frac{1}{q^4} -\frac{2q\cdot p}{q^6} 
+ \cdots \r)G_{n+1}(p,k_1,...,k_2) \quad.
\end{split}
\end{equation}
As usual, the propagators carrying the 
high momenta $q$ can be expanded in powers of \hspace{1cm}
$|p\cdot q||q|^{-2}, m^2|q|^{-2} \ll 1$, which can be 
re-interpreted as the insertion of derivatives and
mass multiplications of the single field $\ft(p)$.
In this case, 
a weaker version of the OPE formula would hold, 
with the Wilson coefficients being simply multiplied
by ${\co}(q\wedge p)$. 
The question is whether or not there is a single operator 
insertion with the net momentum $p$.

This possibility is eliminated when 
the high momentum between the two 
external legs flows through an intermediate leg. 
In this case, we have different ways to share 
the total momentum insertion $p=p_1+p_2$.
Without having to analyze the case where there is a loop 
momentum flowing between $p_1$ and $p_2$
\footnote{which certainly would be a necessary step 
in order to proof an OPE formula.},
we can easily identify problems with a universal
OPE in terms of single local operators with momentum $p$.
If we look at the diagrams of the sort
\begin{equation}
\begin{split}
\hspace{1cm}
&\raisebox{-10ex}{\includegraphics[width=45ex]{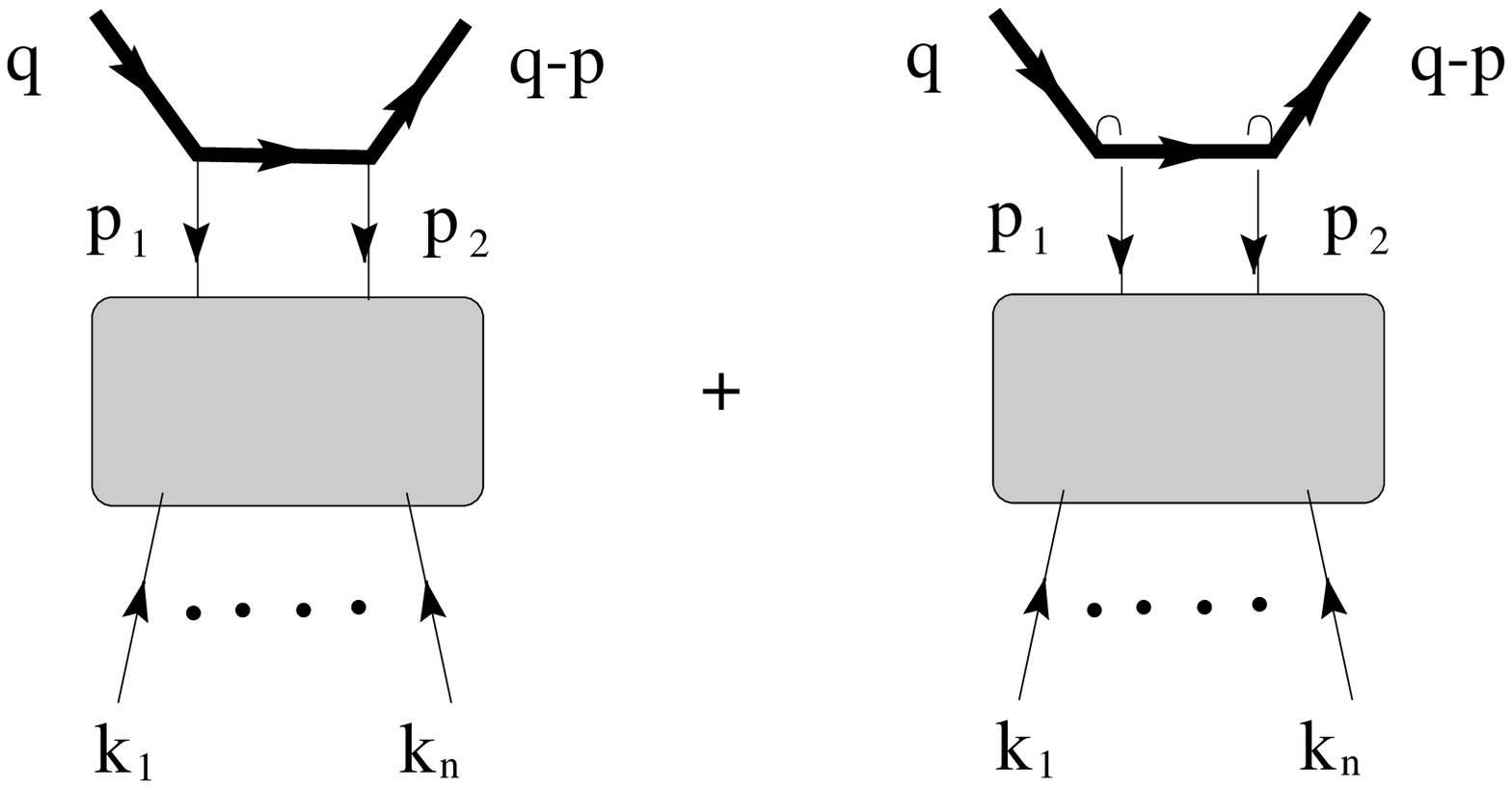}} 
= {\co}(q\wedge p){\co}(p_1\wedge p_2)
\\
&\cdot \l(\frac{1}{q^2+m^2}\r)\l(\frac{1}{(q-p)^2+m^2}\r)
\l(\frac{1}{(q+p_1)^2+m^2}\r)
G_{n+2}(p_1,p_2,k_1,...,k_2) \quad,
\end{split}
\end{equation}
we still have the same global factor ${\co}(q\wedge p)$.
The factor ${\co}(p_1\wedge p_2)$ could be explained 
as coming from the insertion of $\f\star\f$ and 
convenient derivatives of it.
But there are more graphs, the ones given by 
crossing only one low momentum leg with the high-momentum ones:
\begin{equation}
\begin{split}
&\raisebox{-12ex}{\includegraphics[width=45ex]{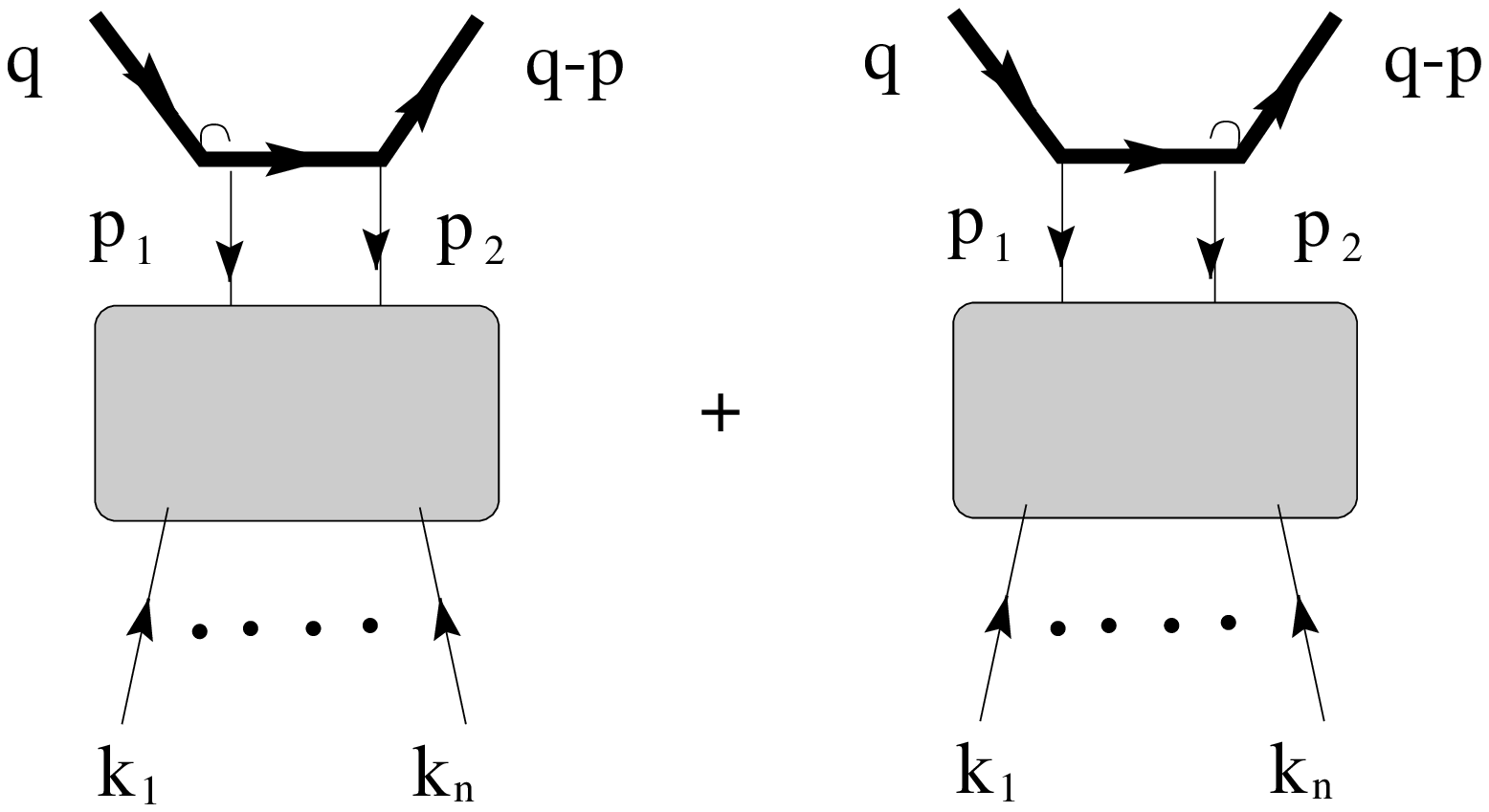}} 
= {\co}(q\wedge (p_2-p_1) +p_1\wedge p_2)
\\
&\cdot\l(\frac{1}{q^2+m^2}\r)\l(\frac{1}{(q-p)^2+m^2}\r)
\l(\frac{1}{(q-p_1)^2+m^2}\r)
G_{n+2}(p_1,p_2,k_1,...,k_2) \quad.
\label{cross}
\end{split}
\end{equation}
From these we see that, due to the 
presence of the arbitrary momentum 
$p_2-p_1$ into the overall phase in \eqref{cross},
there is no way to reproduce it as a series
of insertions of single composite operators 
$\widetilde{\cO}_n(p)$.

\bigskip

\section{Conclusions}

Essentially, there are three results reported in this paper.

In section 2, we concluded that if the renormalization 
program (\`a la Dyson) for $n$-point correlation functions
can be performed at higher loops,
then the same would be valid for single insertions 
of composite operators where all the products are
given by $\star$-products.
We obtained that, as in commutative QFT, 
the composite operators require their own renormalization,
with the possibility of getting mixed.
Then, we have found their insertion at zero
momentum in the noncommutative directions to be 
generically singular.
The reason being that, contrary to the commutative QFT, their 
renormalization (at $\widetilde{p}\not=0$) does not correspond 
to the renormalization of the associated 
coupling parameters.

By the one-loop analysis in section 3,
we saw that the singularity associated with the 
product of two fields in scalar NCQFT is shifted by 
an amount proportional to the momentum flowing 
into the graphs. 
This provides another manifestation 
of the UV/IR mixing in noncommutative field theories.

Finally, an explicit check in momentum space
for the noncommutative $\f^3$ field theory showed   
a breakdown of the OPE as a replacement
of the operator product by a series insertion of
single local operators.

\bigskip
\section*{Acknowledgements}

I thank J. Distler, W. Fischler and E. Gorbatov 
for fruitful discussions.
This work has been supported by the NSF Grant PHY9511632 and
the Robert A.~Welch Foundation.

\bigskip
\section*{Appendix: Useful one-loop integrals}

\begin{equation}
\begin{split}
&I_{D-4}(q;\kt) = \int \frac{d^D p}{(2\pi)^D}\
\frac{e^{ip\kt}}
{\l(p^2 +m^2 \r)\l((p+q)^2 +m^2 \r)} 
\\
&=\int_0^1 \frac{d\a}{8\pi^2}\ e^{-i\a q\kt}  
\l(\frac{\sqrt{q^2\a(1-\a)+m^2}}{2\pi|\kt|}\r)^{\frac{D}{2}-2}
K_{\frac{D}{2}-2}\l[|\kt|\sqrt{q^2\a(1-\a)+m^2}\r] \quad.
\end{split}
\end{equation}
\begin{equation}
\begin{split}
&J_{D-6}(q_1,q_2;\kt) = \int \frac{d^D p}{(2\pi)^D}\
\frac{e^{ip\kt}}
{\l(p^2 +m^2 \r)\l((p+q_1)^2 +m^2 \r)\l((p+q_2)^2 +m^2 \r)} 
\\
&=\frac{1}{32\pi^3}
\int_0^1 d\a_1\int_0^{1-\a_1}d\a_2 \ e^{-i(q_1\a_1+q_2\a_2)\kt}
\Delta^{D-6} \l(2\pi|\kt|\Delta\r)^{3-\frac{D}{2}}
K_{\frac{D}{2}-3}\l[|\kt|\Delta\r] \quad,
\end{split}
\end{equation}
where $\Delta^2=q_1^2\a_1(1-\a_1) +q_2^2\a_2(1-\a_2) +
2q_1q_2\a_1\a_2 +m^2$.

\bigskip

\renewcommand{\baselinestretch}{1} \normalsize
\bibliography{jhep_ope}
\bibliographystyle{JHEP}

\end{document}